\documentclass[a4paper]{report}
\usepackage[utf8]{inputenc}
\usepackage[T1]{fontenc}
\usepackage{RJournal}
\usepackage{amsmath,amssymb,array}
\usepackage{booktabs}

\usepackage{multirow}
\usepackage[ruled,vlined]{algorithm2e}

\def\RR{{\mathbb R}}
\def\vol{\mbox{vol}}

\begin{document}

\sectionhead{Chalkis Apostolos and Vissarion Fisikopoulos}

\begin{article}
\title{\pkg{volesti}: Volume Approximation and Sampling for Convex Polytopes in R}
\author{by Apostolos Chalkis and Vissarion Fisikopoulos}

\maketitle

\abstract{
Sampling from high dimensional distributions and volume approximation of convex bodies are fundamental operations that appear in optimization, finance, engineering, artificial intelligence and machine learning. 
In this paper we present \pkg{volesti}, an {R} package that provides efficient, scalable algorithms for volume estimation, uniform and Gaussian sampling from convex polytopes. 
\pkg{volesti} scales to hundreds of dimensions, handles efficiently three different types of polyhedra and provides non existing sampling routines to  {R}.
We demonstrate the power of \pkg{volesti} by solving several challenging problems using the {R} language.
}

\section{Introduction}\label{sec:intro}

High-dimensional sampling from multivariate distributions with Markov Chain Monte Carlo (MCMC) algorithms is a fundamental problem with many applications in science and engineering~\citep{Iyengar88,Somerville98,Genz09,Schellenberger09,venzke2019}. 
In particular, multivariate integration over a convex set and volume approximation of such sets ---a special case of integration--- have accumulated a broad amount of effort over the last decades. 
Nevertheless, those problems are computationally hard for general dimension~\citep{Dyer88}. 
MCMC algorithms have made remarkable progress solving efficiently the problems of sampling and volume estimation of convex bodies while enjoying great theoretical guarantees~\citep{Chen18,Lee18,Mangoubi19}. However, theoretical algorithms cannot be applied efficiently to real life computations. For example, the asymptotic analysis by~\cite{Lovasz06} hides some large constants in the complexity and in~\citep{Lee18} the step of the random walk used for sampling is too small to be an efficient choice in practice.   
Therefore, practical algorithms have been designed by relaxing the theoretical guarantees and applying new algorithmic and statistical techniques to perform   efficiently while on the same time meet the requirements for high accuracy results~\citep{Emiris14,Cousins15,CoolBod}.

In this paper, we present  \CRANpkg{volesti}~\citep{volesti}, an {R} package containing a variety of high-dimensional MCMC methods for sampling from multivariate distributions restricted to a convex polytope and randomized algorithms for volume estimation of convex polytopes. 
In particular, it includes efficient implementations of three practical volume algorithms---Sequence of Balls (SoB)~\citep{Emiris14}, Cooling Gaussians (CG)~\citep{Cousins15} 
and Cooling convex Bodies (CB)~\citep{CoolBod}. 
In addition to volume estimation, \pkg{volesti} provides efficient implementations for Random-Directions and Coordinate-Directions Hit and Run (RDHR and CDHR)~\citep{smithhnr}, Ball Walk (BaW)~\citep{Hastings70}, Billiard Walk (BiW)~\citep{POLYAK20146123}.
The first three can be used to sample from multivariate uniform or spherical Gaussian distributions (centered at any point), while BiW can be employed, by definition, only for uniform sampling.
On the whole, \pkg{volesti} is the first {R} package that:
\begin{enumerate}  
\item[(a)] performs high dimensional volume estimation, 
\item[(b)] handles efficiently three different types of polyhedra in high dimensions, namely H-polytopes, V-polytopes and Z-polytopes,
\item[(c)] provides---previously absent from {R}---MCMC sampling algorithms for uniform and truncated Gaussian distributions, namely BaW, CDHR and BiW,
\item[(d)] solves some challenging problems in finance, engineering and applied mathematics.
\end{enumerate}

On top of \pkg{volesti} presentation, we illustrate the usage of \pkg{volesti} in the study of convergence of various random walks (e.g. Figure~\ref{fig:sampling_eg}) and accuracy of volume estimation methods. Regarding applications, 
in the last section we illustrate how one can (a) exploit \pkg{volesti} to detect shock events in stock markets following the results by~\cite{Cales18}, (b) evaluate zonotope approximation in engineering~\citep{Kopetzki17} and (c) approximate the number of linear extensions of a partially ordered set which is useful in various applications in artificial intelligence and machine learning. 

To improve the presentation of the current paper, detailed comparisons and benchmarking of R packages--including \pkg{volesti}--for solving the problems of MCMC sampling, volume computation and numerical integration are presented in a separate blog post~\citep{volesti_blog}. 

\subsection{Related {R} software and applications}

Considering MCMC methods to sample from multivariate distributions they are divided in two main categories: truncated to a convex body and untruncated distributions.
For the first category---which clearly is the main focus of this paper---an important case is the truncated Gaussian distribution which arises in several applications in statistics. 
In~\citep{Bolin15} they sample from truncated Gaussian distributions in a novel importance sampling method to study Markov processes that exceed a certain level. 
\cite{Wadsworth14} use sampling from a specific truncated Gaussian distribution to develop a novel method for likelihood inference, while~\cite{Huser13} sample from the same distribution for likelihood estimation for max-stable processes. 
In curve prediction they exploit Gaussian sampling to compute simultaneous  confidence  bands to forecast a full curve from explanatory variables~\citep{Azais10}.  
\cite{Grun12} study the posterior distribution for Bayesian inference
on mixed regression models to represent human immunodeficiency virus
ribonucleic acid levels which is a Gaussian restricted to a convex polytope. 
In~\citep{Albert93} the probit regression model for binary outcomes have an underlying normal regression structure on latent continuous data; sampling from the posterior distribution of the parameters involves sampling from a truncated Gaussian distribution.

Another important special case is the truncated uniform distribution.
In systems biology the flux space of a metabolic network is represented by a convex polytope~\citep{CousinsChnr}; uniform sampling from the interior of that polytope could lead to important biological insights.
In computational finance the set of all possible portfolios in a stock market is in general a convex polytope. 
Volume computation and uniform sampling from that set is useful for crises detection~\citep{Cales18} and efficient portfolio allocation and analysis~\citep{PST04, HHPS2002}.

Considering {R} packages for the truncated case, there is \CRANpkg{tmg}~\citep{tmg} implementing exact Hamiltonian Monte Carlo (HMC) with boundary reflections as well as \CRANpkg{multinomineq}~\citep{multinomineq}, \CRANpkg{lineqGPR}~\citep{lineqGPR}, \CRANpkg{restrictedMVN}~\citep{restrictedMVN}, \CRANpkg{tmvmixnorm}~\citep{tmvmixnorm} implementing variations of the Gibbs sampler.
To our knowledge, the only two {R} packages for uniform sampling is  \CRANpkg{hitandrun}~\citep{hitandrun} and \CRANpkg{limSolve}~\citep{limSolve} which exposes the {R} function \code{xsample()}~\citep{xsample09}. 
For the untruncated case, packages \CRANpkg{HybridMC}~\citep{HybridMC}, \CRANpkg{rhmc}~\citep{rhmc}  and \CRANpkg{mcmc}~\citep{mcmc}, \CRANpkg{MHadaptive}~\citep{MHadaptive} provide implementations for HMC and Metropolis Hastings algorithms respectively. 
For volume computation, the only existing package, \CRANpkg{geometry}~\citep{geometry} computes the volume of  the  convex hull of a set of points and is based on the {C++} library {qhull}~\citep{qhull}.

\section{Algorithms and polytopes}\label{sec:background}

\subsection{Convex polytopes}\label{subsec:polytopes}

\begin{figure}[t]
\centering
\begin{minipage}[h]{0.3\textwidth}
\includegraphics[width=\linewidth]{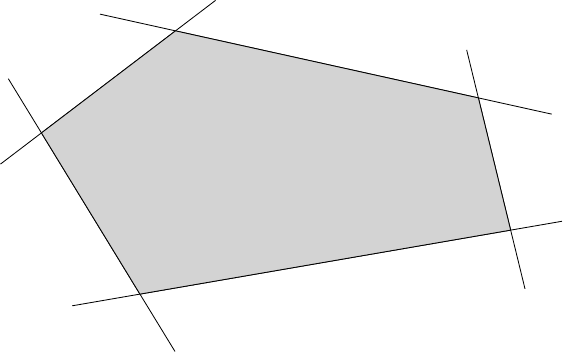}
\end{minipage}
\hspace{.3cm}
\begin{minipage}[h]{0.3\textwidth}
\includegraphics[width=\linewidth]{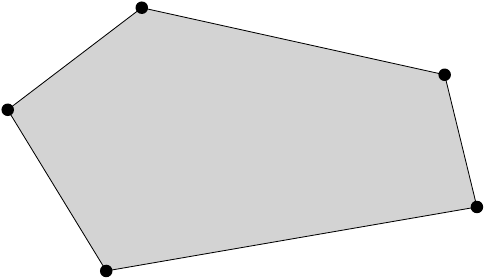}
\end{minipage}
\hspace{.3cm}
\begin{minipage}[h]{0.3\textwidth}
\includegraphics[width=\linewidth]{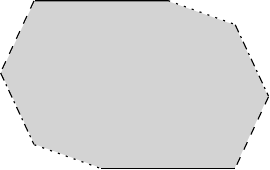}
\end{minipage}
\caption{Examples of three different polytope representations. From left to right: an H-polytope, a V-polytope and a Z-polytope (a sum of four segments). \label{fig:poly_representations}}
\end{figure}

Convex polytopes are a special case of convex bodies with special interest in many scientific fields and applications. For example, in optimization the feasible region of a linear program is a polytope and in finance the set of portfolios is usually expressed by a polytope (i.e.\ the simplex). 
More formally, an H-polytope is defined as 
\[P:=\{ x\ |\ Ax\leq b \}\subseteq\RR^d\] 
where $A\in\RR^{m\times d}$ and $b\in\RR^m$, and we say that $P$ is given in H-representation. Each row $a_i^T\in\RR^d$ of matrix $A$ corresponds to a normal vector that defines a halfspace $a_i^Tx\leq b_i,\ i=[m]$. The intersection of those halfspaces defines the polytope $P$ and the hyperplanes $a_i^Tx=b_i,\ i=[m]$ are called facets of $P$. 
A V-polytope is given by a matrix $V\in\RR^{d\times n}$ which contains $n$ points column-wise, and we say that $P$ is given in V-representation. The points of $P$ that cannot be written as convex combinations of other points of $P$ are called vertices. The polytope $P$ is defined as the convex hull of those vertices, i.e.\ the smallest convex set that contains them. Equivalently, a V-polytope can be seen as the linear map of the canonical simplex $\Delta^{n-1} := \{ x\in\RR^n\ |\ x_i\geq 0,\ \sum_{i=1}^nx_i=1 \}$ according to matrix $V$, i.e.,
\begin{align*}
P:=\{& x\in\RR^d\ |\ \exists y\in \Delta^{n-1} : x=Vy \}
\end{align*}
A Z-polytope (or zonotope) is given by a matrix $G\in\RR^{d\times k}$, which contains $k$ segments column-wise, which are called generators. In this case, $P$ is defined as the Minkowski sum of those segments and we say that it is given in Z-representation. We call \textit{order} of a Z-polytope the ratio between the number of segments over the dimension. Equivalently, $P$ can be expressed as the linear map of the hypercube $[-1,1]^k$ with matrix $G$, i.e.\ 
\[P:=\{ x\in\RR^d\ |\ \exists y\in [-1,1]^k : x=Gy \}.\] 
Thus, a Z-polytope is a centrally symmetric convex body, as a linear map of an other centrally symmetric convex body. 
Examples of an H-polytope, a V-polytope and a Z-polytope in two dimensions are illustrated in Figure~\ref{fig:poly_representations}.
For an excellent introduction to polytope theory we recommend the book of~\citet{Ziegler95}.

\subsection{MCMC sampling and geometric random walks}\label{subsec:sampling}

\begin{figure}[t]
\centering
\begin{minipage}[h]{0.24\textwidth}
\includegraphics[width=\linewidth]{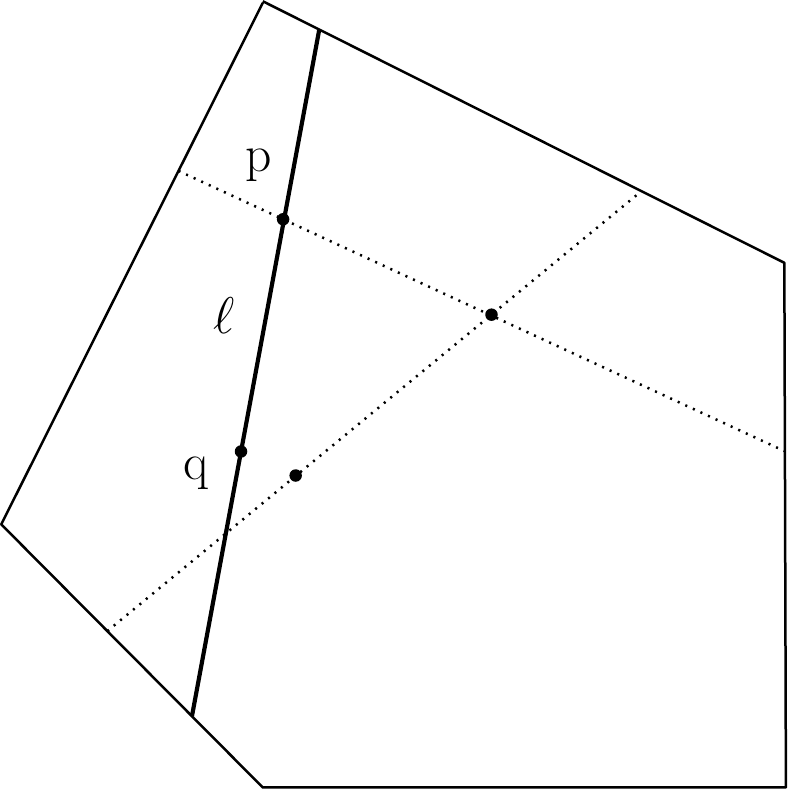}
\end{minipage}
\begin{minipage}[h]{0.24\textwidth}
\includegraphics[width=\linewidth]{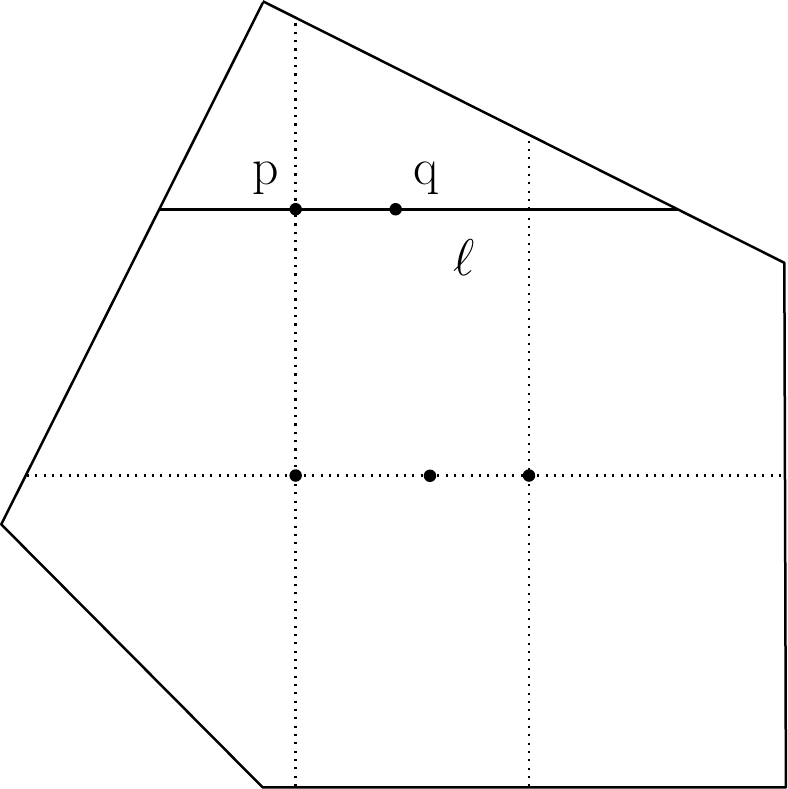}
\end{minipage}
\begin{minipage}[h]{0.24\textwidth}
\includegraphics[width=\linewidth]{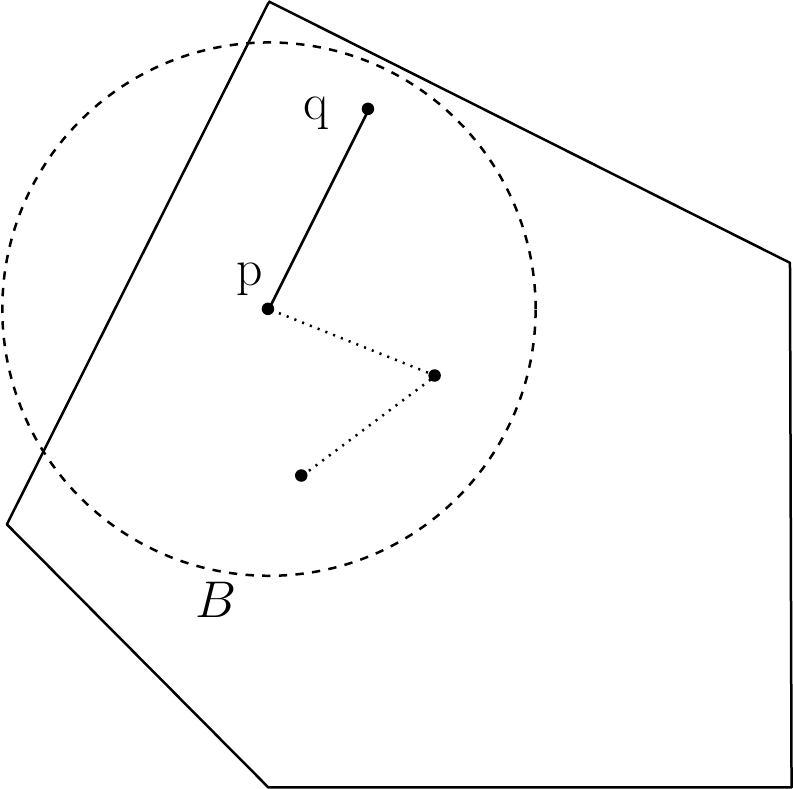}
\end{minipage}
\begin{minipage}[h]{0.24\textwidth}
\includegraphics[width=\linewidth]{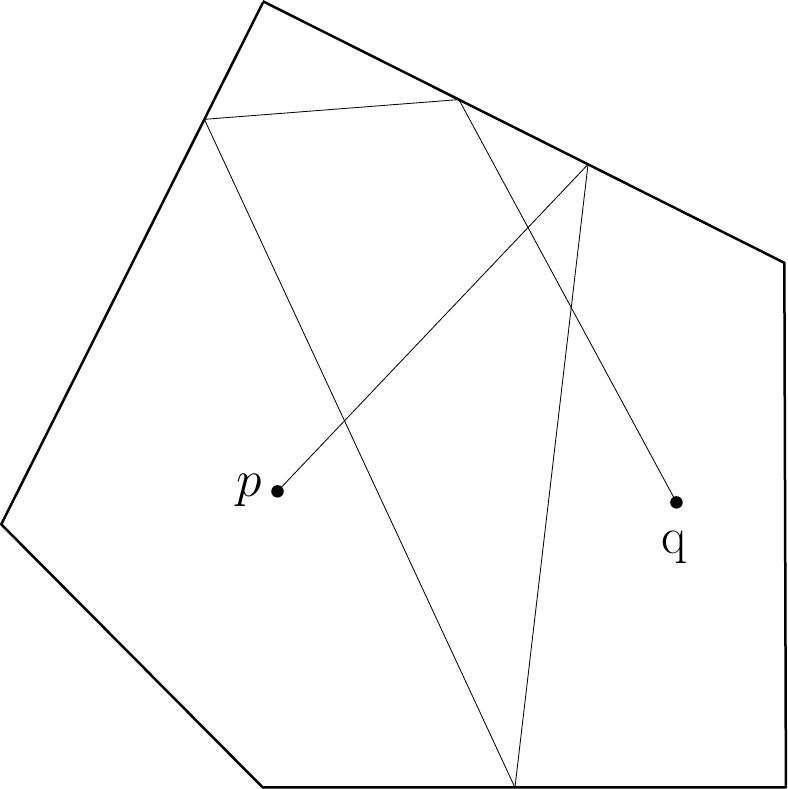}
\end{minipage}
\caption{Examples of random walks. From left to right: RDHR, CDHR, BaW, BiW; $p$ is the point at current step and $q$ the new point computed; $\ell$ is the line computed by RDHR and CDHR; $B$ is the ball computed by BaW. Dotted lines depict previous steps. \label{fig:random_walks}}
\end{figure}

We define here more formally the four geometric random walks implemented in \pkg{volesti}, namely, Hit and Run (two variations, RDHR and CDHR), Ball walk (Baw) and Billiard walk (BiW). They are illustrated in Figure~\ref{fig:random_walks} for two dimensions.

In general if $f:\mathbb{R}^n\rightarrow\mathbb{R}_{+}$ is a non-negative integrable function then it defines a measure $\pi_f$ on any measurable subset $A$ of $\RR^d$,
$$
\pi_f(A)=\frac{\int_A f(x)dx}{\int_{\mathbb{R}^d}f(x)dx}
$$

Let $\ell$ be a line in $\mathbb{R}^d$ and let $\pi_{\ell,f}$ be the restriction of $\pi$ to $\ell$,
$$
\pi_{\ell ,f}(P) = \frac{\int_{p+tu\in P}f(p+tu)dt}{\int_{\ell}f(x)dx}
$$
where $p$ is a point on $\ell$ and $u$ a unit vector parallel to $\ell$. 

Algorithm~\ref{alg:hnr} describes the general Hit and Run procedure.
When the line $\ell$ in line (1.) of the pseudocode is chosen uniformly at random from   all possible lines passing through $p$ then the walk is called Random-Directions Hit and Run~\citep{smithhnr}.
To pick a random direction through point $p\in\mathbb{R}^d$ we could sample $d$ numbers $g_1,\dots ,g_d$ from $\mathcal{N}(0,1)$ and then the vector $u=(g_1,\dots ,g_d)/\sqrt{\sum g_i^2}$ is uniformly distributed on the surface of the $d$-dimensional unit ball. 
A special case is called Coordinate-Directions Hit and Run~\citep{smithhnr} where we pick $\ell$ uniformly at random from the set of $d$  lines that passing through $p$ and are parallel to the 
coordinate axes. 

\begin{algorithm}[H]
  \caption{Hit\_and\_run$(P,p,f)$}
	\label{alg:hnr}
	\SetKwInOut{Input}{Input}
	\SetKwInOut{Output}{Output}
\Input{Polytope $P\subset\RR^d$, point $p\in P$, $f:\RR^d\rightarrow\RR_{+}$}
\Output{A point $q\in P$}
\BlankLine
\begin{enumerate}
\item Pick a line $\ell$ through $p$.
\item \textbf{return} a random point on the chord $\ell \cap P$ chosen from the distribution $\pi_{\ell,f}$.
\end{enumerate}
\end{algorithm}

The Ball walk (Algorithm~\ref{alg:ball}) needs, additionally to Hit and Run, a radius $\delta$ as input. In particular, Ball walk is a special case of Metropolis Hastings~\citep{Hastings70} when the target distribution is truncated.
For both Hit and Run and Ball walk $\pi_f$ is the stationary distribution of the random walk. 
If $f(x)=e^{-||x-x_0||^2/2\sigma^2}$ then the target distribution is the multidimensional spherical Gaussian with variance $\sigma^2$ and its mode at $x_0$. When $f$ is the indicator function of $P$ then the target distribution is the uniform distribution.

\begin{algorithm}[H]
  \caption{Ball\_walk$(P,p,\delta ,f)$}
	\label{alg:ball}
	\SetKwInOut{Input}{Input}
	\SetKwInOut{Output}{Output}
\Input{Polytope $P\subset\RR^d$, point $p\in P$, radius $\delta$, $f \RR^d\rightarrow\RR_{+}$}
\Output{A point $q\in P$}
\BlankLine
\begin{enumerate}
\item Pick a uniform random point $x$ from the ball of radius $\delta$ centered at $p$
\item \textbf{return} $x$ with probability $\min \left\{ 1,\frac{f(x)}{f(p)} \right\}$; \textbf{return} $p$ with the remaining probability.
\end{enumerate}
\end{algorithm}

Billiard walk, is a random walk for sampling from  the uniform distribution~\citep{POLYAK20146123}. It tries to emulate the movement of a gas particle during the physical phenomena of filling uniformly a vessel. Algorithm~\ref{alg:billiard} implements Billiard walk, where  $\langle \cdot , \cdot \rangle$ is the inner product between two vectors, $||\cdot||$ is the $\ell^2$ norm and $|\cdot|$ is the length of a segment.

\begin{algorithm}[H]
  \caption{Billiard\_walk$(P, p, \tau , R)$}
	\label{alg:billiard}
	\SetKwInOut{Input}{Input}
	\SetKwInOut{Output}{Output}
\Input{Polytope $P\subset\RR^d$, current point of the random walk $p\in P$, length of trajectory parameter $\tau \in \RR_{+}$, upper bound on the number of reflections $R \in \mathbb{N}$}
\Output{A point $q\in P$}
\BlankLine
\begin{enumerate}
\item Set the length of the trajectory $L \leftarrow -\tau\ln\eta$, $\eta\sim \mathcal{U}(0,1)$;\\ Set the number of reflections $n\leftarrow 0$ and $p_0\leftarrow p$;\\ Pick a uniformly distributed direction on the unit sphere, $v$;
\item Update $n\leftarrow n+1$; \textbf{If} $n>R$ \textbf{return} $p_0$;
\item Set $\ell \leftarrow \{p + tv, 0\leq t\leq L\}$;
\item \textbf{If} $\partial P\cap\ell=\emptyset$ \textbf{return} $p+Lv$;
\item Update $p \leftarrow \partial P\cap\ell$; Let $s$ be the inner normal vector of the tangent plane on $p$, s.t.\ $||s|| = 1$; Update  $L \leftarrow L - |P\cap\ell|$, $v \leftarrow v - 2\langle v,s \rangle s$; \textbf{goto} 2; 
\end{enumerate}
\end{algorithm}

Every random walk starts from a point in the convex body and perform a number of steps called {\it walk length}. The larger the walk length is the less correlated the final with the starting point will be. 
The number of steps to get an uncorrelated point, that is a point approximately drawn from $\pi_f$, is called {\it mixing time}. We call {\it cost per step} the number of operations performed to generate a point. Hence, the total cost to generate a random point is the mixing time multiplied by the cost per step. 

\begin{table}[h!]
\centering
\begin{tabular}{lccc}
\toprule
\multirow{2}{*}{random walk} & \multirow{2}{*}{mixing time} & cost/step & cost/step \\ 
 & & H-polytope & V- $\&$ Z-polytope \\ \midrule
RDHR \citep{Lovasz06} &  $O^*(d^3)$ & $O(md)$ & $2$ LPs\\
CDHR \citep{Laddha20} &  $O^*(d^{10})$ & $O(m)$ & $2$ LPs\\
BaW \citep{bwmixing} &  $O^*(d^{2.5})$ & $O(md)$ & $1$ LP\\
BiW \citep{POLYAK20146123} & \textbf{?} & $O((d+R) m)$ & $R$ LPs\\
\bottomrule
\end{tabular}
\caption{Overview of the random walks implemented in \pkg{volesti}. LP for linear program; $R$ for the number of reflections per point in BiW; $D$ for the diameter of the polytope.\label{tab:random_walks}}
\end{table}

Table~\ref{tab:random_walks} displays known complexities for mixing time and cost per step. For the mixing time of RDHR we assume that $P$ is well rounded, i.e.\ $B_d\subseteq P\subseteq C\sqrt{d}B_d$, where $B_d$ is the unit ball and $C$ a constant. In general if $rB_d\subseteq P\subseteq RB_d$ then RDHR mixing time is $O^*(d^2(R/r)^2)$. For the mixing time of Ball walk in Table~\ref{tab:random_walks} we assume that $P$ is in isotropic position and the radius of the ball is $\delta = \Theta(1/\sqrt{d})$ \citep{bwmixing}. 
There are no theoretical bounds on mixing time for CDHR and BiW. 
\cite{POLYAK20146123} experimentally shows that BiW converges faster than RDHR when $\tau \approx diam(P)$, i.e.\ the diameter of $P$.
CDHR is the main paradigm for sampling in practice from H-polytopes, e.g.\ in volume computation~\citep{Emiris14} and biology~\citep{CousinsChnr}. The main reason behind this is the small cost per step and the same convergence in practice as RDHR \citep{Emiris14}. For V- and Z-polytopes the cost per step of BiW is comparable with that of CDHR and moreover, converges fast to the uniform distribution~\citep{CoolBod}.
The fact that all above walks are implemented in \pkg{volesti} enable us to empirically evaluate their mixing time using \R\ (e.g., Figure~\ref{fig:sampling_eg}). 

\subsection{Volume estimation}\label{subsec:volume}

As mentioned before, volume computation is a hard problem, so given a polytope $P$ we have to employ randomized algorithms to approximate $\vol(P)$ within some target relative error $\epsilon$ and high probability. 
The keys to success of those algorithms are the Multiphase Monte Carlo (MMC) technique and sampling from multivariate distributions with geometric random walks.

In particular, we define a sequence of functions $\{f_0,\ \dots ,f_q\}$, $f_i:\RR^d\rightarrow \RR$. Then $\vol(P)$ is given by the following telescopic product:
\begin{equation}\label{telegeneric}
\vol(P) = \int_P dx = \int_P f_q(x)dx\frac{\int_P f_{q-1}(x)dx}{\int_P f_q(x)dx}\cdots\frac{\int_P dx}{\int_P f_0(x)dx}
\end{equation}

Then, we need to:
\begin{itemize}
\item Fix the sequence such that $q$ is as small as possible.
\item Select $f_i$ such that each integral ratio can be efficiently estimated.
\item Estimate $\int_P f_q(x)dx$.
\end{itemize}

For a long time researchers, e.g.,~\cite{LovSim}, set $f_i$ to be indicator functions of concentric balls intersecting $P$. It follows that $\int_Pf_i(x)dx = \vol(B_i\cap P)$ and the sequence of convex bodies $P=P_1\supseteq\dots\supseteq P_q,\ P_i = B_i\cap P$ forms a telescopic product of ratios of volumes, while for $\vol(P_q)$ there is a closed formula.
Assuming $rB_d\subset P\subset RB_d$, then $q = O(d\lg R/r)$. The trick now is that we do not have to compute the exact value of each ratio $r_i=\vol(P_i)/\vol(P_{i+1})$, but we can use sampling-rejection to estimate it within some target relative error $\epsilon_i$. If $r_i$ is bounded then $O(1/\epsilon_i^2)$ uniformly distributed points in $P_{i+1}$ suffices. Another crucial aspect is the sandwiching ratio $R/r$ of $P$ which has to be as small as possible. 
This was tackled by a rounding algorithm, that is bringing $P$ to nearly isotropic position~\citep{LovSim}.

The SoB algorithm follows this paradigm and deterministically defines the sequence of $P_i$ such that $0.5\leq\vol(P_{i}) / \vol(P_{i+1})\leq 1$. In the CG algorithm, each $f_i$ is a spherical multidimensional Gaussian distribution and the algorithm uses an annealing schedule~\citep{LovVem} to fix the sequence of those Gaussians.  The SoB algorithm uses a similar annealing schedule but to fix a sequence of convex bodies $P_i$.
As far as performance is concerned, the CB algorithm is the most efficient choice for H-polytopes in less than 200 dimensions and for V- and Z-polytopes in any dimension. For the rest of the cases the user should choose CG algorithm.

\section{Package}\label{sec:package}

The package \pkg{volesti} combines the efficiency of {C++} and the popularity and usability of {R}. 
The package is using the {eigen} library~\citep{eigenweb} for linear algebra, {lpsolve} library~\citep{lpsolve} for solving linear programs and {boost random} library~\citep{boostrandom} (part of Boost {C++} libraries) for random numbers and distributions.
All the code development is performed on \href{https://github.com/GeomScale/volume_approximation}{github platform}.
The package is available in Comprehensive R Archive Network (CRAN) and is regularly updated with new features and bug-fixes.
We employ \href{https://github.com/GeomScale/volume_approximation/actions}{continuous integration} to test the package on various systems and deploy environments.
There is a detailed \href{https://cran.rstudio.com/web/packages/volesti/volesti.pdf}{documentation} of all the exposed \R\ classes and functions publicly available.
We maintain a \href{https://github.com/GeomScale/volume_approximation/blob/develop/CONTRIBUTING.md}{contribution tutorial} to help users and researchers who want to contribute to the development or propose a bug-fix. 
The package is shipped under the \texttt{LGPL-3} license to be open to all the scientific communities.
We use \CRANpkg{Rcpp}~\citep{rcpp} to interface {C++} with {R}.
In particular, we create one \pkg{Rcpp} function for each procedure (such as sampling, volume estimation etc.) and we export it as an {R} function. 

In the following sections we demonstrate the use of  \pkg{volesti}. 
The {R} scripts in the following sections use only standard {R} functions, \pkg{volesti}. In a single script in Section~\ref{sec:applications} we use \CRANpkg{Rfast} to compute the assets' compound return in a stock market.



\subsection{Polytope classes and generators}\label{subsec:generators}

The package \pkg{volesti} comes with three classes to handle different representations of polytopes. 
Table~\ref{tab:classpol} demonstrates the exposed {R} classes. 
The name of the classes are the names of polytope representations as defined in the previous section. 
Each polytope class has a few variable members that describe a specific polytope, demonstrated in Table~\ref{tab:classpol}. 
The matrices and the vectors in Table~\ref{tab:classpol} correspond to those in the  polytope definitions. 
The \code{integer} variable \code{type} implies the representation: \code{1} is for H-polytopes, \code{2} for V-polytopes, \code{3} for Z-polytopes. The \code{numerical} variable \code{volume} corresponds to the  volume of the polytopes if it is known. 
\pkg{volesti} provides standard and random polytope generators. The first, produce well known polytopes such as cubes, cross polytopes, and simplices and assign the value of the exact volume to \code{volume} variable. The second, are random generators using various probability distributions and methods to produce a variety of different random polytopes; notably the generated polytopes has unknown volume. 

\begin{table}[h!]
\centering
\begin{tabular}{ccc}
\toprule
Class & Constructor & Variable members \\ \midrule
{\code{"Hpolytope"}} & {\code{Hpolytope(A,b)}} & $A\in\RR^{m\times d}$, $b\in\RR^m$, \code{integer type}, \code{numerical volume} \\ 
{\code{"Vpolytope"}} & {\code{Vpolytope(V)}} & $V\in\RR^{n\times d}$, \code{integer type}, \code{numerical volume} \\ 
{\code{"Zonotope"}} & {\code{Zonotope(G)}} & $G\in\RR^{k\times d}$, \code{integer type}, \code{numerical volume} \\\bottomrule
\end{tabular}
\caption{\label{tab:classpol} Overview of the polytopes' classes in \pkg{volesti}.}
\end{table}

\subsection{Uniform sampling from polytopes}

A core feature of \pkg{volesti} is approximate sampling from convex bodies with uniform or spherical Gaussian target distribution using the four geometric random walks defined above. 

The following {R} script samples $1000$ points from the $100$-dimensional hypercube $[-1,1]^{100}$ defined as $P$ and stores them in a list. 
 
\begin{example}
R> d = 100
R> P = gen_cube(d, 'H')

R> samples = sample_points(P, random_walk = list(
                           "walk" = "RDHR", "burn-in"=1000, "walk_length" = 5),
                           n = 1000)                              
\end{example}

We use the Random Directions Hit-and-Run (RDHR) walk. 
Other choices are: Coordinate Directions Hit-and-Run (CDHR), Ball Walk (BaW) and Billiard Walk (BiW).
Setting the parameter \code{burn-in} to 1000 means that \pkg{volesti} burns the first 1000 points RDHR generates; setting walk\_length to 5 means that we keep in the list one every five generated points.
The default choice for the target distribution is the uniform distribution.

To evaluate the efficiency of \pkg{volesti} sampling routines one could measure the run-time and estimate the effective sample size~\citep{geyer11} per second. 
To estimate the effective sample size in {R}, a standard choice is the package \CRANpkg{coda}~\citep{coda}. 
In~\citep{volesti_blog} benchmarks show that \pkg{volesti} can be up to $\sim 2\, 500$ times faster than \CRANpkg{hitandrun} for uniform sampling from a polytope.

\begin{figure}[t]
\centering
\includegraphics[width=0.18\textwidth]{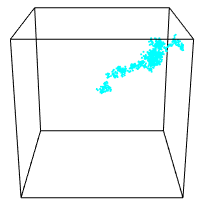}
\includegraphics[width=0.18\textwidth]{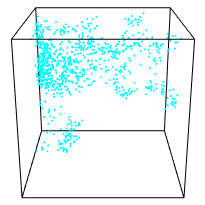}
\includegraphics[width=0.18\textwidth]{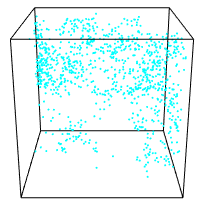}
\includegraphics[width=0.18\textwidth]{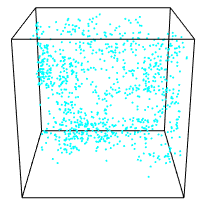}
\includegraphics[width=0.18\textwidth]{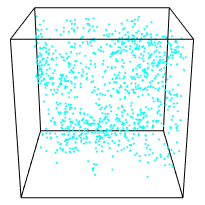}
\includegraphics[width=0.18\textwidth]{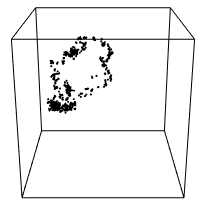}
\includegraphics[width=0.18\textwidth]{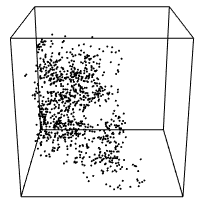}
\includegraphics[width=0.18\textwidth]{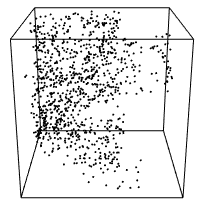}
\includegraphics[width=0.18\textwidth]{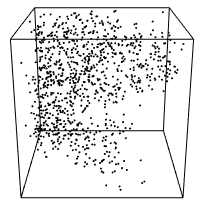}
\includegraphics[width=0.18\textwidth]{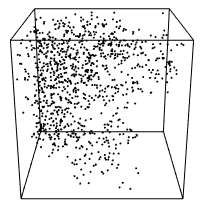}
\includegraphics[width=0.18\textwidth]{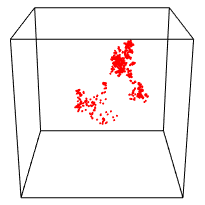}
\includegraphics[width=0.18\textwidth]{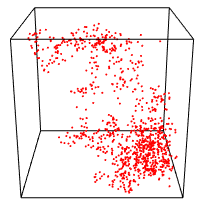}
\includegraphics[width=0.18\textwidth]{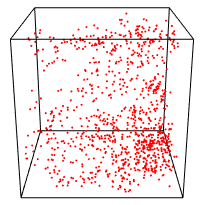}
\includegraphics[width=0.18\textwidth]{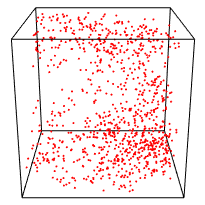}
\includegraphics[width=0.18\textwidth]{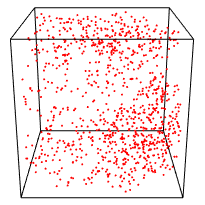}
\includegraphics[width=0.18\textwidth]{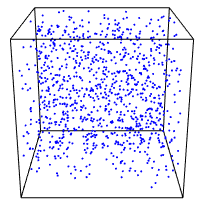}
\includegraphics[width=0.18\textwidth]{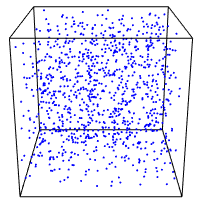}
\includegraphics[width=0.18\textwidth]{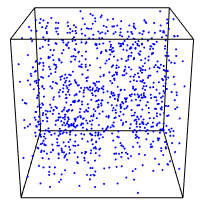}
\includegraphics[width=0.18\textwidth]{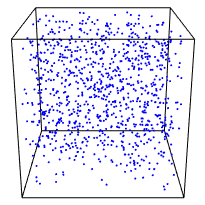}
\includegraphics[width=0.18\textwidth]{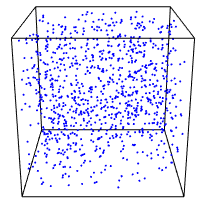}
\caption{Uniform sampling from a random rotation of the hypercube $[-1,1]^{200}$. We map the sample back to $[-1,1]^{200}$ and then we project them on $\RR^3$ by keeping the first three coordinates. Each row corresponds to a different walk: BaW, CDHR, RDHR, BiW. Each column to a different walk length: \{1, 50, 100, 150, 200\}. That is, the sub-figure in third row and forth column corresponds to RDHR with 150 walk length.
\label{fig:sampling_eg}}
\end{figure}

Moreover, using \pkg{volesti} and {R} we can empirically study the mixing time of the geometric random walks implemented in \pkg{volesti}. 
To this end, we sample uniformly from a random rotation of the 200-dimensional hypercube $[-1,1]^{200}$. 
First, we generate the hypercube and use \code{rotate\_polytope()} that returns the rotated polytope and the matrix of the linear transformation.

\begin{example}
R> d = 200
R> num_of_points = 1000
R> P = gen_cube(d, 'H')
R> retList = rotate_polytope(P, rotation = list("seed" = 5))
R> T = retList$T
R> P = retList$P
\end{example}

Then, we use \code{sample\_points()} to sample from the rotated cube with various walk lengths to test the practical mixing of the random walk. 

\begin{example}
R> for (i in c(1, seq(from = 50, to = 200, by = 50))){
     points1 = t(T) 
                                      "walk" = "BaW", "walk_length" = i, "seed" = 5))
     points2 = t(T) 
                                      "walk" = "CDHR", "walk_length" = i, "seed" = 5))
     points3 = t(T) 
                                      "walk" = "RDHR", "walk_length" = i, "seed" = 5))
     points4 = t(T) 
                                      "walk" = "BiW", "walk_length" = i, "seed" = 5))
   }
\end{example}

Finally, we map the points back to $[-1,1]^{200}$ using the inverse transformation and then we project all the sample points on $\RR^3$, or equivalently on the 3D cube $[-1,1]^3$, by keeping the first three coordinates. We plot the results in Figure~\ref{fig:sampling_eg}.

Note that, in general, perfect uniform sampling in the rotated polytope would result to perfect uniformly distributed points in the 3D cube $[-1,1]^3$. 
Hence, Figure~\ref{fig:sampling_eg} shows an advantage of BiW in mixing time for this scenario compared to the other walks---it mixes relatively well even with one step (i.e.\ walk length). 
Notice also that the mixing of both CDHR and RDHR seem similar while it is sightly better than the mixing of BaW. 

\subsubsection{Gaussian sampling from polytopes}

In many Bayesian models the posterior distribution is a multivariate Gaussian distribution restricted to a specific domain. We illustrate the usage of \pkg{volesti} for the case of the truncation being the canonical simplex $\Delta^n =\{ x\in\R^n\ |\ x_i\geq 0,\ \sum_ix_i=1 \}$, which is of special interest. This situation typically occurs whenever the unknown parameters can be interpreted as fractions or probabilities. Thus, it appears in many important applications~\citep{Altmann14}.
In particular, we consider the following density,
\begin{equation}\label{eq:gaussian_simplex}
f(x|\mu,\Sigma) \propto  \left\{
\begin{array}{ll}
      exp[-\frac{1}{2}(x-\mu)^T\Sigma(x-\mu)],  & \mbox{ if } x\in\Delta^n ,\\
      0, & \mbox{otherwise.}\\
\end{array} 
\right.  
\end{equation}

Clearly, the support of the density in Equation~(\ref{eq:gaussian_simplex}) is defined by a convex subset of a linear subspace of $\RR^n$. Thus, to sample from $f(x|\mu,\Sigma)$ we apply a proper linear transformation, induced by a matrix $N\in\RR^{n\times (n-1)}$, that maps the support to a full dimensional polytope in $\RR^{n-1}$, while the covariance matrix changes accordingly to $\Sigma' = N^T\Sigma N$. Then, we apply a Cholesky decomposition to $\Sigma'=LL^T$ and employ the linear transformation induced by $L$ to transform the distribution into a spherical Gaussian distribution. 

In the following {R} script we we first generate a random 100-dimensional positive definite matrix $\Sigma$. Then, we sample from the multivariate Gaussian distribution with the covariance
matrix being $\Sigma$ and the mode being the center of the canonical simplex $\Delta^n$. To achieve this goal we first 
apply all the necessary linear transformations to both the probability density function and the $\Delta^n$ to obtain the standard Gaussian distribution,  $\mathcal{N}(0,I_n)$, restricted to a general full dimensional simplex. 

\begin{example}
R> d = 100
R> S = matrix( rnorm(d*d,mean=0,sd=1), d, d) #random covariance matrix 
R> S = S 
R> shift = rep(1/d, d)
R> A = -diag(d)
R> b = rep(0,d)
R> b = b - A 
R> Aeq = t(as.matrix(rep(1,d), 10,1))
R> N = pracma::nullspace(Aeq)       
R> A = A 
R> S = t(N) 
R> A = A 
R> P = Hpolytope(A=A, b=as.numeric(b)) #new truncation
\end{example}

Next, we use the \code{sample\_points()} function to sample from the standard Gaussian distribution restricted to the computed simplex and we apply the inverse transformations to obtain a sample in the initial space.

\begin{example}
R> samples = sample_points(P, n = 100000, random_walk = 
                           list("walk"="CDHR", "burn-in"=1000, 
                           "starting_point" = rep(0, d-1),
                           distribution = list("density" = "gaussian", 
                           "mode" = rep(0, d-1)))) 
R> samples_initial_space = N 
   kronecker(matrix(1, 1, 100000), matrix(shift, ncol = 1))
\end{example}

In the previous script we set the starting point of the walk to the mode of the Gaussian i.e.\ the origin. 
Note that the default choice in \pkg{volesti} for the target distribution in the case of Gaussian sampling is the standard Gaussian; that is the target distribution in the above script. 

Considering comparisons, \pkg{volesti} is at least one order of magnitude faster than \CRANpkg{restrictedMVN} and \pkg{tmg} for computing a sample of similar quality.
For more details on comparison with other packages we refer to~\citep{volesti_blog}. 

\subsection{Volume estimation}\label{subsec:volume_volesti}

Let us now give an example of how we approximate the volume of a polytope in \pkg{volesti}.
Since this is a randomized algorithm it makes sense to compute some statistics for the output values using {R} when approximating the volume of the $10$-dimensional cube $[-1,1]^{10}$ generated as an H-polytope.

\begin{example}
R> P = gen_cube(10, 'H')
R> volumes = list()
R> for (i in seq_len(20)) {
     volumes[[i]] = volume(P, settings = list("error" = 0.2))
   }
\end{example}

\begin{figure}[t]
\centering
 \begin{minipage}[h]{0.495\textwidth}
\includegraphics[width=\linewidth]{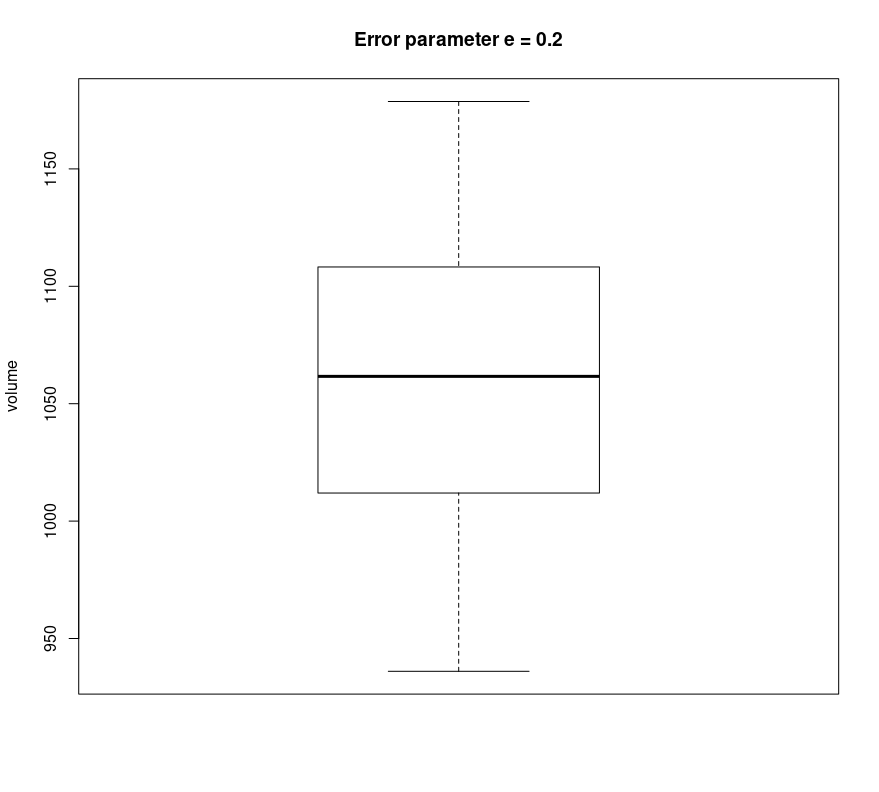}
\end{minipage}
\begin{minipage}[h]{0.495\textwidth}
\includegraphics[width=\linewidth]{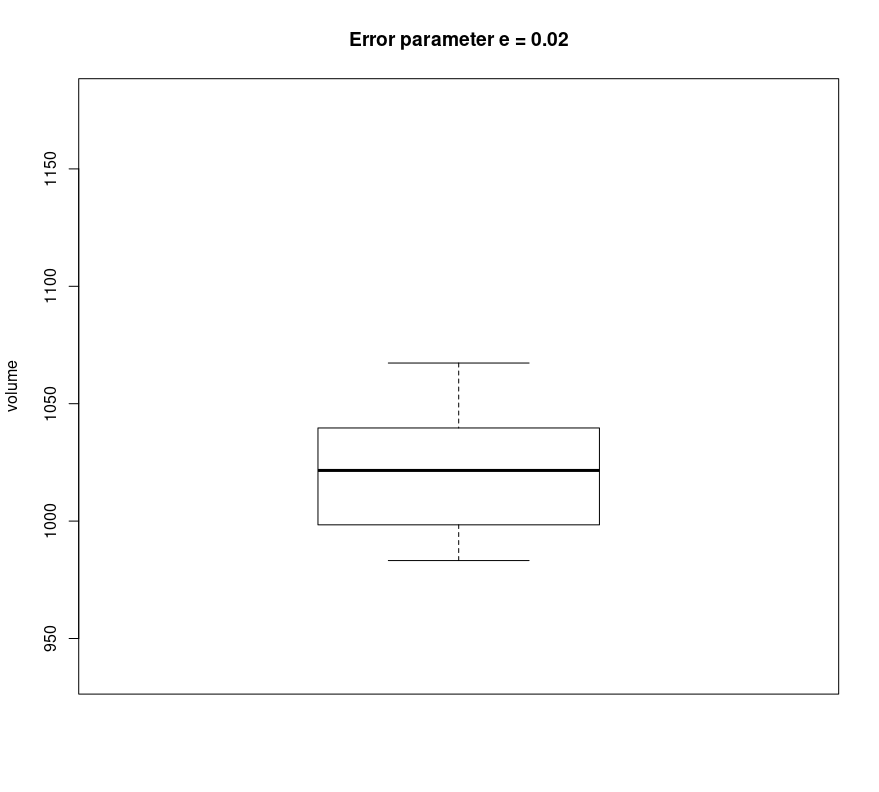}
\end{minipage}
\caption{The boxplot of the estimated volumes of the hypercude $[-1,1]^{10}$ by \pkg{volesti}. Left the input error parameter is $\epsilon=0.2$ and right is $\epsilon=0.02$. \label{fig:avg_volume}}
\end{figure}

By changing the error to $0.02$ we can obtain more accurate results.
The results are illustrated in Figure~\ref{fig:avg_volume}.
Note that the exact volume is $1024$.

To understand the need of randomized computation in high dimensions implemented in \pkg{volesti} we can consider the state-of-the-art volume computation in {R} today namely, \CRANpkg{geometry}. It implements a deterministic algorithm which run-time grows exponentially with the dimension. Because of the later property, \CRANpkg{geometry} in general, fails to terminate for polytope in dimension $d\geq 20$.
See~\citep{volesti_blog} for comparison details with \CRANpkg{geometry}.

The following script illustrates the usage and efficiency of \pkg{volesti} to compute the volume of high-dimensional polytopes. 
In particular, a V-polytope, namely the \href{https://en.wikipedia.org/wiki/Cross-polytope}{cross-polytope}, and an H-polytope, namely the hypercube.

\begin{example}
R> d = 80
R> P = gen_cross(80, 'V')  #generate a cross polytope in V-representation

R> time = system.time({
          volume_estimation = volume(P, settings = list(
                                     "algorithm" = "CB", "random_walk" = "BiW", 
                                     "seed" = 127)) })
R> exact_volume = 2^d/prod(1:d)
R> cat(time[1], abs(volume_estimation - exact_volume) / exact_volume)

82.874   0.074434

R> P = gen_cube(d, 'H')   #generate a hypercube polytope in H-representation

R> time = system.time({
          volume_estimation = volume(P, settings = list(
                                     "algorithm" = "CB", "random_walk" = "CDHR", 
                                     "seed" = 23)) })
R> exact_volume = 2^d
R> cat(time[1], abs(volume_estimation - exact_volume) / exact_volume)

0.657   0.067633
\end{example}

For V- and Z- polytopes the most efficient choice of random walk is BiW, while for H-polytopes is CDHR; this explains why we use different random walks in the previous script. 
However, notice that the run-time for the H-polytope is two order of magnitude smaller. This happens because the cost per step of a random walk in a V-polytope increases comparing to H-polytopes. 

Last but not least, \pkg{volesti} provides random polytope generators. 
The following command estimates the volume of a randomly generated V-polytope 
that is the convex hull of $40$ uniformly generated random points from the $20$-dimensional cube.  
\begin{example}
R> P = gen_rand_vpoly(20, 40, generator = list("body" =  "cube", "seed" = 1729))
R> volume_estimation = volume(P)
\end{example}

The next call estimates the volume of an H-polytope randomly generated  
as an intersection of $180$ linear halfspaces computed by random tangent hyperplanes on an $60$-dimensional hypersphere. 

\begin{example}
R> P = gen_rand_hpoly(60, 180, generator = list('constants' = 'sphere'))
R> volume_estimation = volume(P)
\end{example}

Since the exact volume of those polytopes is unknown, the accuracy of the computed estimation is unknown and statistical methods such as the effective sample size~\citep{geyer11} could be used.

\section{Applications}\label{sec:applications}
We demonstrate \pkg{volesti}'s potential to solve challenging problems. More specifically, we provide detailed use-cases for applications in finance (crises detection and portfolio scoring), decision and control, multivariate integration and artificial inteligence. 

\subsection{Financial crises detection and portfolio scoring}\label{subsec:fintools}

In this subsection we present how one could employ \pkg{volesti} to detect financial crises or shock events in stock markets by following the method of~\cite{Cales18}.
For all the examples in the sequel we use a set of $52$ popular exchange traded funds (ETFs) and the US central bank (FED) rate of return publicly available from \url{https://stanford.edu/class/ee103/portfolio.html}. The following script is used to load the data.

\begin{example}
R> MatReturns = read.table("https://stanford.edu/class/ee103/data/returns.txt", 
                           sep = ",")
R> MatReturns = MatReturns[-c(1, 2), ]
R> dates = as.character(MatReturns$V1)
R> MatReturns = as.matrix(MatReturns[ ,-c(1, 54)])
R> MatReturns = matrix(as.numeric(MatReturns), nrow = dim(MatReturns )[1], ncol =
                       dim(MatReturns )[2], byrow = FALSE)
R> nassets = dim(MatReturns)[2]
\end{example}

\begin{figure}[t]
\centering
 \begin{minipage}[h]{0.35\textwidth}
\includegraphics[width=\linewidth]{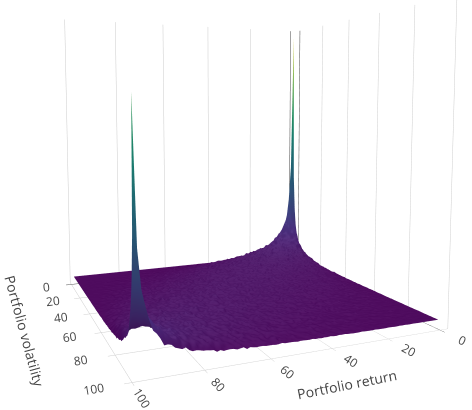}
\end{minipage}
\hspace{1.0cm}
\begin{minipage}[h]{0.35\textwidth}
\includegraphics[width=\linewidth]{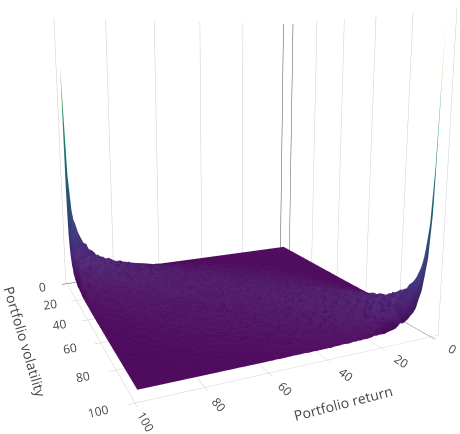}
\end{minipage}
\caption{Left, a copula that corresponds to normal period ($07/03/2007 - 31/05/2007$), $I=0.2316412$. Right, a copula that corresponds to a crisis period ($18/12/2008 - 13/03/2009$), $I=5.610785$; $x$ axis is for return and $y$ axis is for volatility.\label{fig:cop2}}
\end{figure}

The method uses the copula representation to capture the dependence between portfolios' return and volatility. A copula is an approximation of the bivariate joint distribution while both marginals follow the uniform distribution.  In normal times, portfolios are characterized by slightly positive returns and a moderate volatility, in up-market times (typically bubbles) by high returns and low volatility, and during financial crises by strongly negative returns and high volatility. Thus, when a copula implies a positive dependence (see Figure~\ref{fig:cop2} left) then it probably comes from a normal period. On the other side, when the dependence between portfolios' return and volatility is negative (see Figure~\ref{fig:cop2} right), the copula comes probably from a crisis period. The first case occurs when the indicator that computes the ratio between the red mass over the blue mass is smaller than $1$ and the second case when that indicator is larger than $1$. The function \code{copula()} can be used to compute such copulas. When two vectors of returns are given as input by the user, then the computed copula is related to the problem of momentum effect in stock markets.

The following script produces Figure~\ref{fig:cop2} by setting the starting and the stopping date for the left and the right plot respectively. To compute the copula we use the compound asset return which is the rate of return for capital over a cumulative series of time~\citep{Cales18}.

\begin{example}
R> row1 = which(dates 
R> row2 = which(dates 
R> compound_asset_return = Rfast::colprods(1 + MatReturns[row1:row2, ]) - 1  
R> mass = copula(r1 = compound_asset_return, sigma = cov(MatReturns[row1:row2, ]), 
                 m = 100, n = 1e+06, seed = 5)
\end{example}

\begin{figure}[t]
\centering
\includegraphics[width=\linewidth]{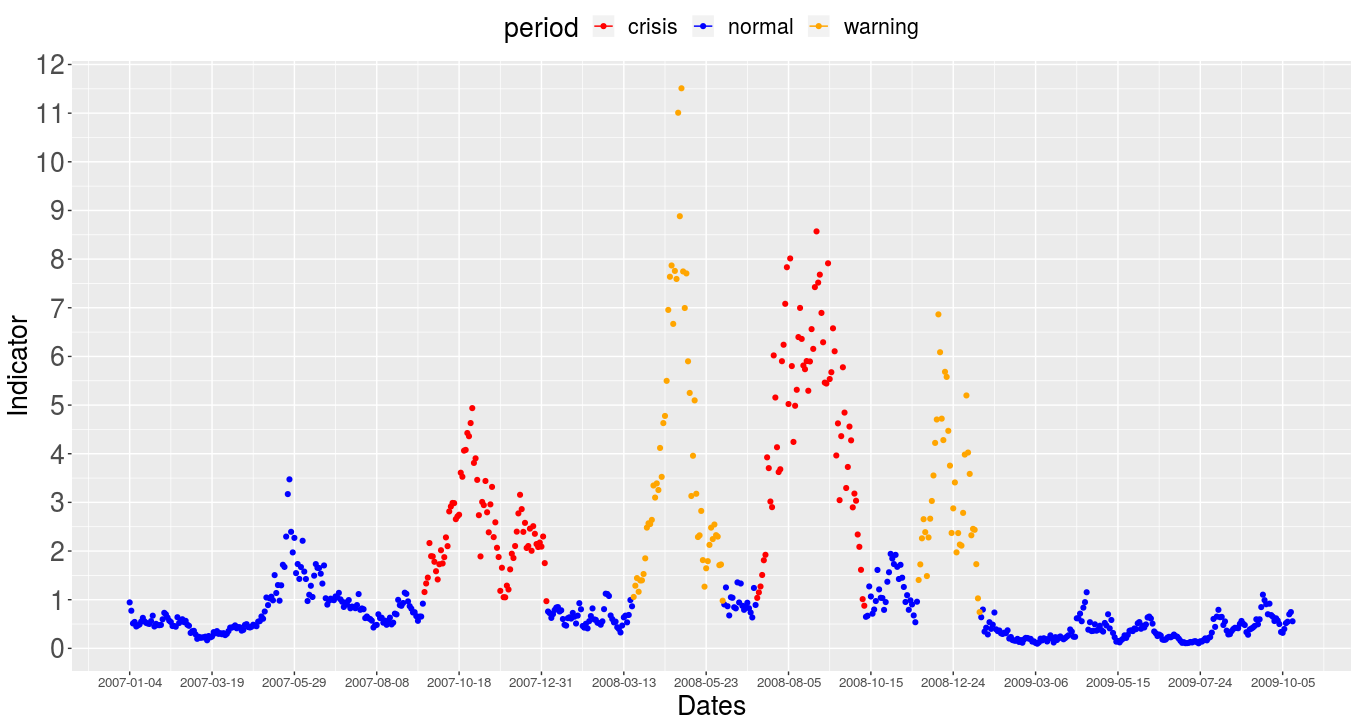}
\caption{The values of the indicators from 2007-01-04 until 2010-01-04. We mark with red the crisis periods, with orange the warning periods and with blue the normal periods that \pkg{volesti} identifies. \label{fig:indicators}}
\end{figure}

Moreover, the function \code{compute\_indicators()} computes the copulas of all the sets of \code{win\_len} consecutive days and returns the corresponding indicators and the states of the market during the given time period. 
The next script takes as input the daily returns of all the $52$ assets from $01/04/2007$ until $04/01/2010$. When the indicator is $\geq 1$ for more than $30$ days we issue a warning and when it is for more than $60$ days we mark this period as a crisis (see Figure~\ref{fig:indicators}).

\begin{example}
R> row1 = which(dates 
R> row2 = which(dates 
R> market_analysis = compute_indicators(returns =  MatReturns[row1:row2, ],
                                        parameters = list("win_len" = 60, "m" = 100,
                                        "n" = 1e+06, "nwarning" = 30, "ncrisis" = 60, 
                                        "seed" = 5))                            
R> I = market_analysis$indicators
R> market_states = market_analysis$market_states
\end{example}

We compare the results with the database for financial crises in European countries proposed in~\citep{ESRB17}. The only listed crisis for this period is the sub-prime crisis (from December 2007 to June 2009). Notice that Figure~\ref{fig:indicators} successfully points out $4$ crisis events in that period ($2$ crisis and $2$ warning periods) and detects sub-prime crisis as a W-shape crisis.

As a second financial application we will use \pkg{volesti} to evaluate the performance of a given portfolio.
In particular, \pkg{volesti} computes the proportion of all possible allocations that the given portfolio outperforms. This score independently introduced in~\citep{Pouchkarev05, Guegan11, Banerjee11} and is an alternative to more classical choices for the evaluation of the performance of a portfolio as the Sharpe-like ratios proposed in the 1960's by  \cite{Jensen67,Sharpe66,Treynor15}. However the efficient computation of that score was uncertain until~\cite{Cales18} notice that Varsi's algorithm~\citep{Varsi73} can be used to perform robust computations in high dimensions. Varsi's algorithm is implemented in \pkg{volesti} by the function \code{frustum\_of\_simplex()} and computes volumes in thousands of dimensions in just a few milliseconds on modest hardware. As an example, the following {R} script, let us know that in $03/13/2009$ any portfolio with return $0.002$ outperforms almost the $48\%$ of all possible portfolios.

\begin{example}
R> R = MatReturns[which(dates 
R> R0 = 0.002
R> tim = system.time({ exact_score = frustum_of_simplex(R, R0) })
R> cat(exact_score, tim[3])

0.4773961  0.001
\end{example}
 
\subsection{Zonotope volumes in decision and control}\label{subsec:eval_zonotopes}

Volume approximation for Z-polytopes (or zonotopes) could be very useful in several applications in decision and control~\citep{Kopetzki17}, in autonomous driving~\citep{autdriv} or human-robot collaboration~\citep{robcol}. The complexity of algorithms that manipulate Z-polytopes strongly depends on their order. Thus, to achieve efficient computations the approach that is common in practice is to over-approximate the Z-polytope at hand $P$, as tight as possible, with a second Z-polytope $P_{red}$ of smaller order. Then the ratio of fitness $\rho = (\vol(P_{red}) / \vol(P))^{1/d}$ is a
good measure for the quality of the approximation.
However, this ratio cannot computed for dimensions typically larger than $10$ (see~\citep{Kopetzki17}).  
\pkg{volesti} is the first software to the best of our knowledge that efficiently approximates the ratio of fitness  of a high dimensional  Z-polytope--typically up to 100 and order 200--or a Z-polytope of very high order in lower dimensions--e.g., order 1500 in 10 dimensions.

\begin{figure}[t]
\centering
\includegraphics[width=0.5\textwidth]{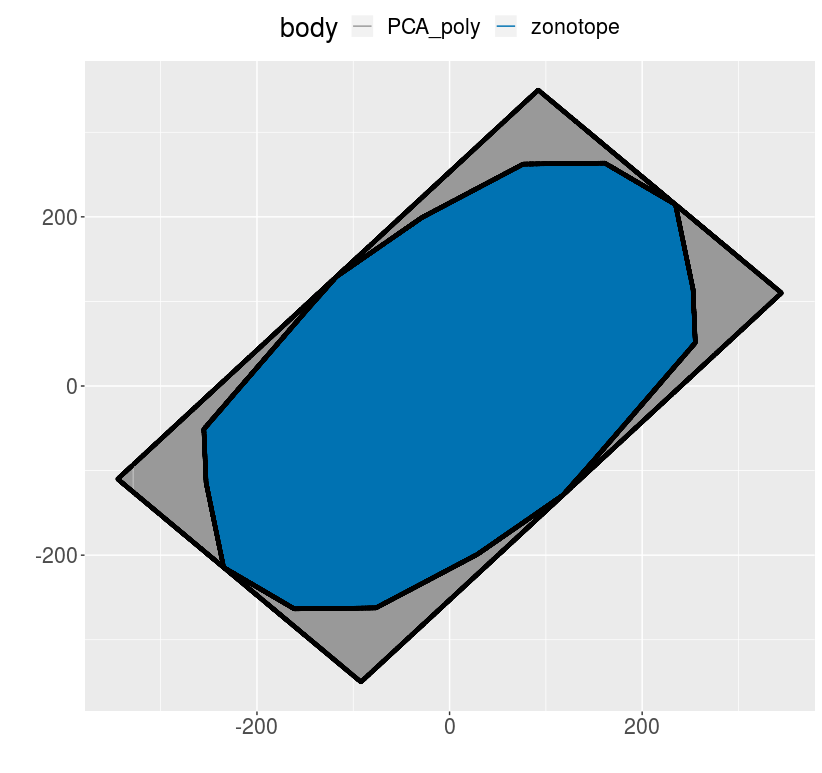}
\caption{With blue color a 2D Z-polytope. With grey color the over-approximation of $P$ computed with PCA method.\label{fig:zono_approx}}
\end{figure}

As an illustration, the following {R} script, generates a random 2D zonotope, computes the over-approximation with the PCA method and estimates the ratio of fitness. The \code{sample\_points} function is then used to plot the two polygons (Figure~\ref{fig:zono_approx}). 

\begin{example}
R> Z = gen_rand_zonotope(2, 8, generator = list("distribution" = "uniform", 
                         "seed" = 1729))
R> points1 = sample_points(Z, random_walk = list("walk" = "BRDHR"), n = 10000)
R> retList = zonotope_approximation(Z = Z, fit_ratio = TRUE, 
                                    generator = list("seed" = 5))
R> P = retList$P
R> cat(retList$fit_ratio)

1.116799539

R> points2 = sample_points(P, random_walk = list("walk" = "BRDHR", "seed" = 5), 
                           n = 10000)
\end{example}

\subsection{High-dimensional integration}\label{subsec:integrals}

Computing the integral of a function over a convex set (i.e.\ convex polytope) is a hard fundamental problem with numerous applications. 
%
\pkg{volesti} can be used to approximate the value of such an integral by a simple MCMC integration method, which employs the $\vol(P)$ and a uniform sample in $P$. In particular, let 
\begin{equation}
I = \int_P f(x)dx .
\end{equation}
Then sample $N$ uniformly distributed points $x_1,\dots ,x_N$ from $P$ and,
\begin{equation}
I\approx \vol(P)\frac{1}{N}\sum_{i=1}^N f(x_i) .
\end{equation}

The following {R} script generates a V-polytope for $d=5,10,15,20$ and estimates the integral of,
\begin{equation}\label{eq:f_int}
	f(\mathbf{x}) = \sum_{i=1}^n x_i + 2x_1^2 + x_2 + x_3,
\end{equation}
over the generated V-polytope $P$. 

Considering the efficiency of \pkg{volesti}, the Table~\ref{tab:integration} reports the exact value of $I$ computed by \CRANpkg{SimplicialCubature}~\citep{SimplicialCubature}. It computes multivariate integrals over simplices. 
Hence, to compute an integral of a function over a convex polytope $P$ in {R} one should compute the Delaunay triangulation with package \CRANpkg{geometry} and then use the package \CRANpkg{SimplicialCubature} to sum the values of all the integrals over the simplices computed by the triangulation. 
The pattern is similar to volume computation, for $d=5,10$ the exact computation is faster than the approximate, for $d=15$ \pkg{volesti} is $13$ times faster and for $d=20$ the exact approach halts while \pkg{volesti} returns an estimation in less than a minute. 

\begin{example}
R> num_of_points = 5000
R> f = function(x) { sum(x^2) + (2 * x[1]^2 + x[2] + x[3]) }
R> for (d in seq(from = 5, to = 20, by = 5)) {
     P = gen_rand_vpoly(d, 2 * d, generator = list("seed" = 127))
       
     points = sample_points(P, random_walk = list("walk" = "BiW", 
                            "walk_length" = 1, "seed" = 5), n = num_of_points)
     sum_f = 0
     for (i in seq_len(num_of_points)){
       sum_f = sum_f + f(points[, i])
     }
     V = volume(P, settings = list("error" = 0.05, "seed" = 5))
     I2 = (sum_f * V) / num_of_points 
   }

\end{example}

\begin{table}[t]
\centering
\begin{tabular}{|c|ccccc|}
\toprule
dimension & \textit{Exact value} & \textit{Estimated value} & \textit{Rel.\ error} & \textit{Exact Time (sec)} & \textit{Est.\ Time (sec)} \\ \hline
5 &  0.02738404  &  0.02446581  &  0.1065667  & 0.023 &   3.983 \\ 
10 & 3.224286e-06 & 3.204522e-06 & 0.00612976 & 3.562  &  11.95 \\
15 & 4.504834e-11 & 4.867341e-11 & 0.08047068 & 471.479 & 33.256 \\
20 & -   &      1.140189e-16 & -   &        -       & 64.058 \\ \hline
\end{tabular}
\caption{\label{tab:integration} We compute the integral of the function in Equation~(\ref{eq:f_int}) over a random generated V-polytope. \textit{Exact value} the exact value of the integral using \CRANpkg{SimplicialCubature} and \CRANpkg{geometry}; \textit{Estimated value} the estimation of the integral with \pkg{volesti}; \textit{Rel.\ error} the relative error of \pkg{volesti}; \textit{Exact Time} the sum of run-times of \CRANpkg{geometry} and \CRANpkg{SimplicialCubature}; \textit{Est.\ Time} the run-time of \pkg{volesti}; "-" indicates that the program halts.}
\end{table}

\subsection{Combinatorics and artificial intelligence}\label{subsec:linear_extensions}
 
We focus now on a different problem, namely, counting the linear extensions of a given partially ordered set (poset), which arises in various applications in artificial intelligence and machine learning such as partial order plans~\citep{Muise16}, and learning graphical models~\citep{Niinim16}.

Let $G= (V, E)$ be an acyclic digraph with $V= [n] :=\{1,2, \dots,n\}$. One might want to consider $G$ as a representation of the poset $V:i > j$ if and only if there is a directed path from node $i$ to node $j$. A permutation $\pi$ of $[n]$ is called a linear extension of $G$ (or the associated poset $V$) if $\pi^{-1}(i)> \pi^{-1}(j)$ for every edge $i\rightarrow j \in E$. 

Let $P_{LE}(G)$ be the polytope in $\RR^n$ defined by 
\[P_{LE}(G) =\{x\in \RR^n\ |\ 1\geq x_i \geq 0 \text{ for all }i=1,2,\dots ,n\},\] and $x_i\geq x_j$ for all directed edges $i\rightarrow j \in E$.
It is well known~\citep{Stanley86} that the number of linear extensions of $G$ equals the normalized volume of $P_{LE}(G)$ i.e., 
\[\#_{LE}G=\vol(P_{LE}(G))\ n!\]

It is also well known that counting linear extensions is \#P-complete~\citep{Brightwell91}. Thus, as the number of graph nodes (i.e., the dimension of $P_{LE}(G)$) grows the problem becomes intractable for exact methods. Interestingly, \pkg{volesti} provides an efficient approximation method that could be added to the ones surveyed by~\cite{Talvitie18}. 

\begin{figure}[t]
\centering
\includegraphics[width=.25\linewidth]{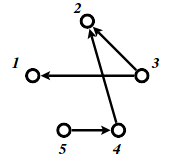}
\caption{An acyclic directed graph with 5 nodes, 4 edges and 9 linear extensions.\label{fig:graph}}
\end{figure}

As a simple example consider the graph in Figure~\ref{fig:graph} that has 9 linear extensions\footnote{Example taken from \url{https://people.inf.ethz.ch/fukudak/lect/pclect/notes2016/expoly_order.pdf}}. 
This number can be estimated in milliseconds using \pkg{volesti} as in the following script, where the estimated number of linear extensions is $9.014706$.


\begin{example}
R> A = matrix(c(
              -1,0,1,0,0,0,
              -1,1,0,0,0,-1,
              0,1,0,0,0,0,-1,
              1,1,0,0,0,0,0,
              1,0,0,0,0,0,1,
              0,0,0,0,0,1,0,
              0,0,0,0,1,-1,
              0,0,0,0,0,-1,
              0,0,0,0,0,-1,
              0,0,0,0,0,-1,
              0,0,0,0,0,-1),
              ncol = 5, nrow = 14, byrow = TRUE)
R> b = c(0, 0, 0, 0, 0, 0, 0, 0, 0, 1, 1 , 1, 1, 1)
R> P_LE = Hpolytope(A = A, b = b)
R> time = system.time({ LE = volume(P_LE, settings = list("error" = 0.01, 
                                    "seed" = 1927)) * factorial(5) })
\end{example}

\section{Concluding remarks and future work}

\pkg{volesti} is an {R} package that provides MCMC sampling routines for multivariate distributions restricted to convex polytopes and volume estimation. It supports three different polytope representations and thus, it is useful for several applications. We illustrate the usage of \pkg{volesti} with simple reproducible examples and show how \pkg{volesti} can be used to address challenging problems in modern applications. 

Regarding future work, the expansion of \pkg{volesti} to support general log-concave sampling methods would be of special interest for several applications. Efficient log-concave sampling could also lead to additional sophisticated methods to estimate a multivariate integral over a convex polytope~\citep{Lovasz06b}.

\section{Computational details}

The results in this paper were obtained using {R} 3.4.4 and {R} 3.6.3 and \pkg{volesti} 1.1.2-2. The versions of the imported by \pkg{volesti} packages are \CRANpkg{stats} 3.4.4~\citep{stats} and \CRANpkg{methods} 3.4.4~\citep{methods}; of the linked by \pkg{volesti} packages, \CRANpkg{Rcpp} 1.0.3, \CRANpkg{BH} 1.69.0.1~\citep{BH}, \CRANpkg{RcppEigen} 0.3.3.7.0~\citep{RcppEigen}, and the suggested package \CRANpkg{testthat} 2.0.1~\citep{testthat}. For comparison to \pkg{volesti} and for plots this paper uses, \CRANpkg{geometry} 0.4.5, \CRANpkg{hitandrun} 0.5.5, \CRANpkg{SimplicialCubature} 1.2, \CRANpkg{Rfast} 2.0.3~\citep{Rfastpkg}. \CRANpkg{ggplot2} 3.1.0~\citep{ggplot2}, \CRANpkg{plotly} 4.8.0~\citep{plotly}, \CRANpkg{rgl} 0.100.50~\citep{rgl}, \CRANpkg{coda} 0.19.4. All packages used are available from \href{http://CRAN.R-project.org}{CRAN}.

All computations were performed on a PC with {\tt Intel\textregistered\ Pentium(R) CPU G4400 @ 3.30GHz $\times$ 2 CPU} and {\tt 16GB RAM}.

\section{Acknowledgments}
The main part of the work has been done while A.C. was supported by \href{https://summerofcode.withgoogle.com}{Google Summer of Code} 2018 and 2019 grants and V.F. was his mentor. 
The authors acknowledge fruitful discussions with Ioannis Emiris, Ludovic Cal\`es, Elias Tsigaridas, the {R}-project for statistical computing and the {R}  community.  

\bibliography{volesti}

\address{Apostolos Chalkis\\
  Department of Informatics \& Telecommunications\\
  National \& Kapodistrian University of Athens\\
  Greece\\
  GeomScale org.\\
  ORCiD: 0000-0002-4628-1907\\
  \email{achalkis@di.uoa.gr}} 

\address{Vissarion Fisikopoulos\\
  Department of Informatics \& Telecommunications\\
  National \& Kapodistrian University of Athens\\
  Greece\\
  GeomScale org.\\
  ORCiD: 0000-0002-0780-666X\\
  \email{vfisikop@di.uoa.gr}}

\end{article}

\end{document}